\documentclass[12pt]{article}

\usepackage[margin=1in]{geometry}
\usepackage{amsmath,bm,amsthm,amssymb,bm}
\usepackage{graphicx}
\usepackage{natbib}
\usepackage{url,hyperref}
\DeclareUnicodeCharacter{00B0}{\ensuremath{{}^0}}
\usepackage{upquote}
\usepackage{color}

\def\shalf{\mbox{{\footnotesize$\frac{1}{2}$}}}
\def\onebym{\mbox{{\small$\frac{1}{m}$}}}
\def\onebytwom{\mbox{{\small$\frac{1}{2m}$}}}
\def\fhalf{\mbox{{\footnotesize$1/2$}}}

\newcommand{\half}{\frac{1}{2}}

\renewcommand{\Bigl}{\Big}
\renewcommand{\Bigr}{\Big}

\newcommand{\Tr}{\mathrm{Tr}}

\newcommand{\var}{\mathrm{var}~}

\def\shalf{\mbox{{\tiny$\frac{1}{2}$}}}
\def\half{\mbox{{\small$\frac{1}{2}$}}}

\newcommand{\bbeta}{{\beta}}

\newcommand{\degc}{{\textsuperscript{o}C}}
\renewcommand{\deg}{{\textsuperscript{o}~}}

\title{On the usefulness of lattice approximations for fractional Gaussian fields}
\author{Somak Dutta\footnote{Address: 2438 Osborn Dr. Ames IA 50011, Email: somakd@iastate.edu} \\ Iowa State University, Ames, IA. \\ Debashis Mondal\\ Oregon State University, Corvallis, OR.}
\date{}

\begin{document}

\maketitle
\begin{abstract}
Fractional Gaussian fields provide a rich class of spatial models and have a long history of applications in multiple branches of science. However, estimation and inference for fractional Gaussian fields present significant challenges. This book chapter investigates the use of the fractional Laplacian differencing on regular lattices to approximate to continuum fractional Gaussian fields. Emphasis is given on model based geostatistics and likelihood based computations. For a certain range of the fractional parameter, we demonstrate that there is considerable agreement between the continuum models and their lattice approximations. For that range, the parameter estimates and inferences about the continuum fractional Gaussian fields can be derived from the lattice approximations. Interestingly, regular lattice approximations facilitate fast matrix-free computations and enable anisotropic representations.  We illustrate the usefulness of lattice approximations via simulation studies and by analyzing sea surface temperature on the Indian Ocean.
\end{abstract}
\paragraph{Keywords} Argo floats; Discrete cosine transformation;  Fractional Laplacian differencing; Geometric anisotropy; H-likelihood; Long-range dependence; MLE; Power-law variogram; Regular lattice;

\section{Introduction}

Fractional Gaussian fields have inspired extensive research in spatial statistics. Fractional fields generalize the notion of fractional noise in two or higher dimensions and are particularly important for studying power laws and modeling long-range dependencies. The early mathematical development of fractional fields can be traced to the works of \citet{yagl:1957}, \citet{whit:1962}, \citet{mcke:1963}, \citet{gang:1986}, \citet{mand:vann:1968},  and others.  Also notable are the works by \citet{dobr:1979}, \citet{yagl:1987}, \cite{gran:joye:1980}, \cite{hosk:1981}, \cite{gay:heyd:1990}, \citet{bera:1994},  \citet{ma:2003} and \citet{kelb:leon:2005}.  Recent surveys on the topic are provided in \citet{chil:delf:2009}, \citet{cohe:ista:2013} and \citet{lodh:shef:2016}.  Fractional Gaussian fields cover, as special cases, the de Wijs process or the Gaussian free fields \citep{math:1970,shef:2007,mond:2015}, the thin plate spline \citep{gu:wahb:1993}, higher-order intrinsic random fields \citep{math:1970,math:1973} and power variogram models. Fractional Gaussian fields can also be seen as limiting cases of the Mat\'ern models.  Their applications range from agriculture, hydrology and environmental science to cosmology, statistical physics, and quantum mechanics.

Advances in fractional Gaussian fields have been accompanied by the development of their discrete-space approximations.  In one dimension, the discrete-space approximations emerged in the influential works of \citet{gran:joye:1980} and  \citet{hosk:1981} on fractional differencing and have received extensive treatments in time series analysis.  Furthermore, there is an impressive array of works on intrinsic autoregressions that can be understood as discrete-space approximations of various intrinsic random fields; see e.g. \citet{kuns:1987}, \citet{besa:koop:1995}, \citet{besa:mond:2005}, \citet{rue:held:2005}, \citet{lind:rue:2011}, \citet{cres:2015}, and \citet{mond:2018}.  In a recent paper, \citet{dutt:mond:2015,dutt:mond:2016b} consider fraction Laplacian differencing as ways to approximate fractional Gaussian fields in two dimensions.   These discrete-space approximations do not conflict, but rather establish a deeper connection with limiting, continuum fractional Gaussian fields, and help  advance statistical computation.

The intent of this book chapter is to provide a basic introduction to fractional Gaussian fields with an emphasis on their interpretation, their statistical properties and on exploring their discrete-space approximations.  We start with a basic definition of  fractional Gaussian fields  in Sections \ref{sec:matern}, which arise when a fractional order of the Laplacian is applied to the Gaussian white noise  on the two dimensional Euclidean space.  We then present their spectral densities, variograms and, in Section \ref{sec:fracdiff}, consider their discrete-space approximations. These discrete-space approximations are obtained by restricting the random fields on regular lattices and  by replacing the Laplacian operator on the two  dimensional Euclidean space with discrete Laplacians on regular grids. We primarily focus on  Gaussian fields with geometric anisotropies which occur when variogram contours are formed by concentric ellipses, and standard statistical analysis presents further challenges. 

In Section \ref{sec:MLestimation} we focus on a certain range of the fractional parameter and discuss maximum likelihood estimation for spatial models based on fractional Gaussian fields or their discrete-space approximations.   We  judge the effectiveness of the discrete-space approximations  in terms of efficiency in approximating the continuum limits.  For this, we consider simulation studies.  Using computer experiments, we demonstrate that discrete-space approximations provide as good estimates  as the limiting, continuum model  based on the fractional Gaussian fields. We further demonstrate statistical scalability. In Section \ref{sec:data}, we present an analysis of the Indian Ocean surface temperature obtained from the Argo floats devices, further establishing the agreement between the models based on continuum fractional Gaussian fields and their discrete-space approximations.  For ease of understanding and reproducibility, we provide all Matlab codes in the appendix.  The dataset can be obtained from 
\url{ftp://usgodae.org/pub/outgoing/argo/geo/indian_ocean/} and also from the corresponding author.

\section{Fractional Gaussian fields and their approximations}
\subsection{Fractional Gaussian fields}\label{sec:matern}

We follow \citet{lodh:shef:2016} and consider anisotropic  two dimensional fractional Gaussian fields as
\begin{equation}\label{eqn:imaternspde}
 \psi(u,v) = ( - \nabla)^{-\nu/2} \xi(u,v), \quad (u,v)\in\mathbb{R}^2,
\end{equation}
where $\xi(u,v)$, for $(u,v)\in\mathbb{R}^2$ represent Gaussian white noise  with marginal variance $\sigma^2$,  $\nu$ denotes the fractional or the long range dependence parameter, and  $\nabla$  is the anisotropic  Laplacian
\begin{equation}\label{eqn:spdeoperator}
\nabla  = 4\alpha\frac{\partial^2 }{\partial u^2} + 4(\half-\alpha)\frac{\partial^2 }{\partial v^2}. 
\end{equation}
The parameter $0 < \alpha <1/2$ controls the degree of geometric anisotropy in both the $x$ and $y$ directions and the value $\alpha= 1/4$ corresponds to isotropic random fields.

It is important to note that a Gaussian white noise is not pointwise defined, rather, it is a generalized random field \citep{math:1973,chil:delf:2009,lodh:shef:2016}. In fact, a white noise $\xi$ on $\mathbb{R}^2$ is defined such that for any pair of disjoint measurable sets $A$ and $B$, $\int_A\xi(u,v)dudv$ and $\int_B\xi(u,v)dudv$ are independent Gaussian random variables with zero means and variances $\sigma^2|A|$ and $\sigma^2|B|$ respectively, where $|A|$ and $|B|$, respectively, denote the areas of $A$ and $B.$ However, either pointwise or in a distributional sense, fractional Gaussian fields exist for all real values for $\nu$. Fractional Gaussian fields include many important models as special cases. In particular, $\nu=0$ corresponds to the White noise model, $\nu=1$ gives the de Wijs process or the Gaussian free fields,  $\nu=2$ indicates thin plate splines (also known as bi-Laplacian random fields) and $\nu =3/2$ denotes the L\'evy Brownian motion. For all $\nu \le 1$, fractional Gaussian fields correspond to generalized random fields and are defined in a distributional sense. For $1  < \nu < 2$, fractional Gaussian fields have stationary (zeroth-order) increments. For $2  \le \nu < 3$, fractional Gaussian fields have
stationary first-order increments, and so on. For non-negative $\nu$, it can be shown that the generalized spectral density of the fractional Gaussian fields in \eqref{eqn:imaternspde} is 
\begin{equation}\label{eqn:sdimatern}
 \rho(\omega,\eta) = \frac{\sigma^2}{\left(4\alpha\omega^2 + 4(\half-\alpha)\eta^2\right)^\nu},~(\omega,\eta) \in \mathbb{R}^2.
\end{equation} 
Thus, for $1  < \nu < 2$,  standard Fourier integral formulas give an expression for variogram  of $\Psi$ as \citep{dutt:mond:2016a}
\begin{eqnarray}\label{eqn:variogPower}
 \gamma(h,k) & = &\half \var(\psi(u+h,v+k)-\psi(u,v)) 
  = \int_{\mathbb{R}^2}\{1-\cos(h\omega+k\eta) \} \rho(\omega,\eta)d\omega d\eta\nonumber \\
 & = & \dfrac{\sigma^2\Gamma(\nu-\shalf)}{16\sqrt{\pi\alpha(\half-\alpha)}\Gamma(\nu)\Gamma(2\nu-1)\sin(-\nu\pi)} \left(\frac{h^2}{4\alpha}+ \frac{k^2}{4(\half-\alpha)}\right)^{\nu-1},
\end{eqnarray}
for any $(h,k)\in\mathbb{R}^2.$ By virtue of \eqref{eqn:variogPower}, fractional Gaussian fields, for values of $1< \nu <2$, correspond to widely used power variogram models in geostatistics.  Fractional Gaussian fields can also be seen as a limiting case of Mat\'ern models. The latter emerge as  a solution to the stochastic partial differential equation 
\begin{equation}\label{eqn:maternspde}
 (\kappa^2 - \nabla)^{\nu/2} \psi^\dagger(u,v) = \xi(u,v),~(u,v)\in\mathbb{R}^2,
\end{equation}
where $\kappa > 0$ is the inverse range parameter.  The limiting cases, as $\kappa \to 0$, provide the  fractional Gaussian fields in \eqref{eqn:imaternspde}.

The spectral density \eqref{eqn:sdimatern} and the variogram function \eqref{eqn:variogPower} play an important role in all subsequent  statistical computations. For example, for $1 < \nu <2$, the variogram function \eqref{eqn:variogPower} is key to computing the actual likelihood function, which we shall discuss in Section \ref{sec:MLiMatern}.

\subsection{Lattice approximations}\label{sec:fracdiff}
For $m\geq 1$, let $\mathbb{Z}^2_m$ denote the sub-lattice of the two-dimensional integer lattice $\mathbb{Z}^2$ with spacing $1/m$. Following \citet{dutt:mond:2016a}, let  $\Delta_m$  be the Laplace difference  operator  on the sub-lattice $\mathbb{Z}^2_m$. Thus,  for any real valued function $w$ defined at the lattice points of $\mathbb{Z}_m^2$, we get 
\[
\Delta_m w(u,v) =  w(u,v) - [ \alpha_m \{ w(u+\onebym ,v) + w(u-\onebym,v)\}  + (\half - \alpha_m) \{ w(u,v+\onebym)+w(u,v-\onebym) \}] ,
\]
where $0 \leq \alpha_m \leq 1/2$. Next, we consider 
\begin{equation}\label{eqn:fld}
 \psi^{(m)}(u,v) = \Delta_m^{- \nu/2}\; \xi^{(m)}_{u,v}, \hspace{.1in} \nu \ge 0,
\end{equation}
where  $\xi^{(m)}_{u,v}$ is a Gaussian white noise on the sub-lattice ${\mathbb Z}_m^2$ with 
\[
\var \xi^{(m)}_{u,v} = \sigma^2_m /m^2.
\]
Then, the  random field $\{ \psi^{(m)}(u,v)\}$ that arises from the above fractional Laplacian differencing can be interpreted as an approximation of the fractional Gaussian random fields \ref{eqn:imaternspde} on the sub-lattice ${\mathbb Z}_m^2$.  It then follows from the standard theory on linear transformation or spectral representation that the generalized spectral density function 
of $\{ \psi^{(m)}(u,v)\}$ has the form 
\begin{equation}\label{eqn:sdffd}
\rho_{m} (\omega,\eta) = \frac{\sigma_m^2}{ m^2  \Bigl[ 4\alpha_m\sin^2(\onebytwom \omega) + 4(\shalf-\alpha_m)\sin^2(\onebytwom \eta)\Bigr]^\nu}, 
\end{equation}
with $\omega, \eta \in (-\pi m, \pi m]$,   $\sigma_m >0$ and  $\nu > 0$. 
Under appropriate scaling of the parameters $\sigma^2_m$  the lattice random field converges to the fractional Gaussian fields  \eqref{eqn:imaternspde}. That is,  as $m\to\infty,$ 
\[
4^\nu m^{2\nu-2}\sigma^2_m \to \sigma^2,
\]
and $\alpha_m \to \alpha,$ the spectral density $\rho_{m}$ converges to  $\rho$  pointwise and in $L_p$ for all $p \leq 2/\lfloor\nu-1\rfloor.$
\begin{figure}
 \centering\includegraphics[width=\textwidth]{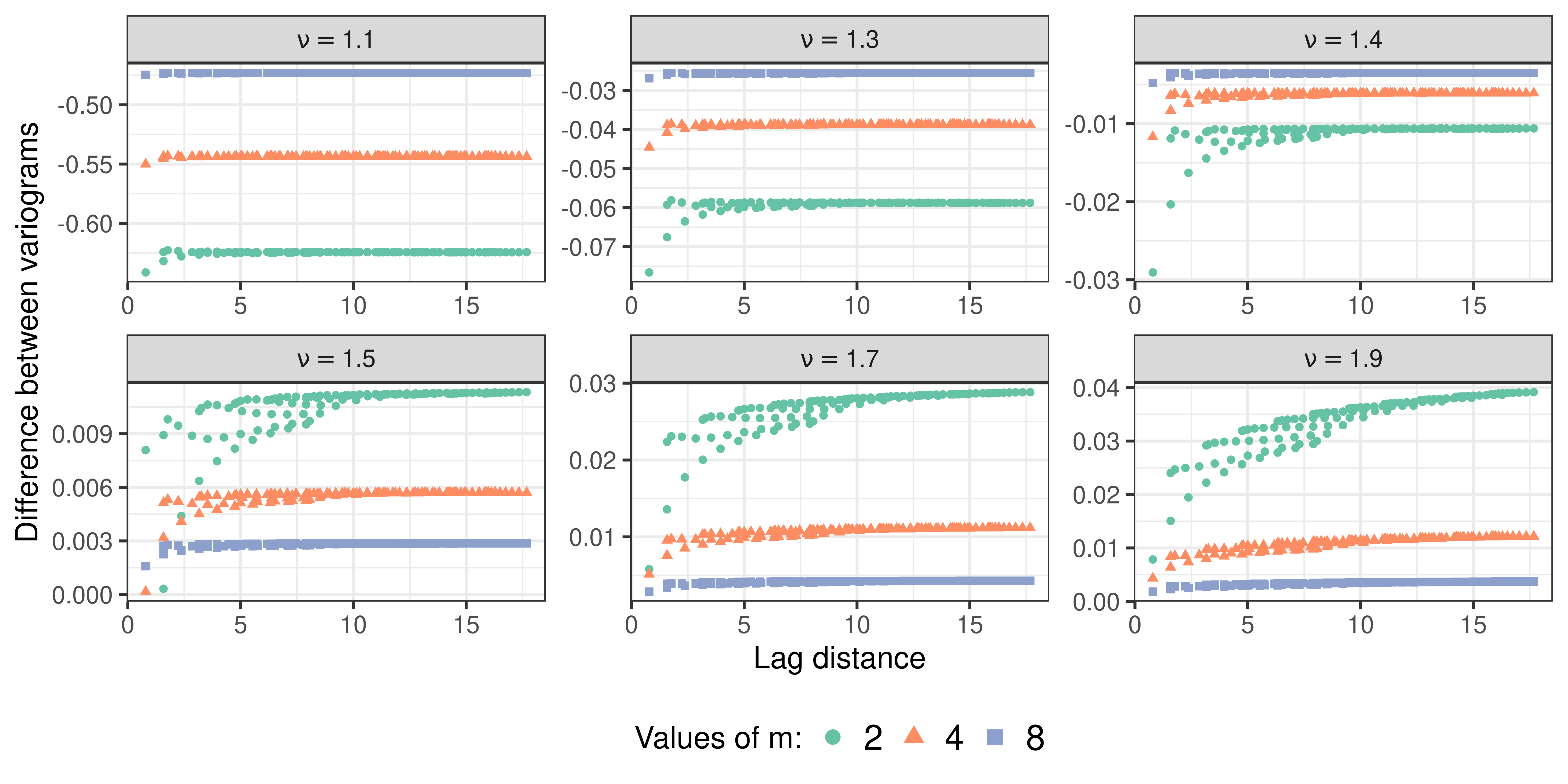}
 \caption{Plot of $\gamma_{m}(h,k)-\gamma(h,k)$ against the lag distance $\sqrt{h^2/\bigl(4\alpha\bigr)+k^2/\bigl(4(\half-\alpha)\bigr)}$ for difference values of $\nu$ and $m.$}
 \label{fig:variodiff}
\end{figure}
The preceding result indicates that that continuum fractional Gaussian fields are scaling limits of fractionally differenced Gaussian random fields on regular lattices and it also explicitly describes the rescaling of parameters needed.

For integer values $\nu=1,2, \ldots$, fractional Laplacian differencing corresponds to intrinsic autoregressions of order $\nu-1$ on the sub-lattice ${\mathbb Z}_m^2$. Furthermore, for $1  < \nu < 2$, fractional  Laplacian differencing leads to a  random field with stationary (zeroth-order) increments. Similarly, for $2  \le \nu < 3$, fractional  Laplacian differencing gives rise to a  random field with stationary first-order increments, and so on. 

For $1 < \nu <2$,  the variogram function of $\psi^{(m)}$ takes the form of 
\begin{eqnarray}\label{eqn:variogsddif}
 \gamma_{m}(h,k) & = &\half\var\big(\psi^{(m)}(u+h,v+k)-\psi^{(m)}(u,v)\big) \nonumber \\
  &= &\frac{1}{4\pi^2}\int_{-m\pi}^{m\pi}\int_{-m\pi}^{m\pi}\{1-\cos(\omega h+\eta k)\} \rho_{m}(\omega, \eta) d\omega d\eta.
\end{eqnarray}
for $(h,k)\in \mathbb{Z}^2$, and one can show that $\sup_{(h,k)\in\mathbb{Z}^2}\left|\gamma_{m}(h,k) - \gamma(h,k)\right| \to 0$ as $m\to \infty$. However, unlike \eqref{eqn:variogPower}, there is no such exact analytic formula available for \eqref{eqn:variogsddif}.  Interestingly, we can apply the numerical method presented in \citet{dutt:mond:2016b} to calculate \eqref{eqn:variogsddif} and assess  how well $\gamma_{m}$ in \eqref{eqn:variogsddif} approximate the limiting variogram function \eqref{eqn:variogPower}. The plots in Figure \ref{fig:variodiff} display the difference $\gamma_{m}(h,k)-\gamma(h,k)$ for $\sigma^2= 1,$ $\sigma^2_m = 4^{-\nu}m^{2-2\nu},$ $\alpha_m \equiv \alpha=0.1$ and  various values of $\nu$ between $1$ and $2$ and, for $m=2,4$ and $8.$  We find that the difference is essentially a small constant independent of the spatial lag, but depending on $\nu$ and $\alpha.$ Because the variogram of a nugget effect is constant,  these results thus suggest that, when augmented with a nugget effect, the fractionally differenced random field at the one-eighth lattice provides an excellent approximation of an fractional Gaussian field plus a nugget effect on the original lattice. This  approximation result is consistent with the isotropic case discussed in \citet{dutt:mond:2016b}.


\section{Model based geostatistics}

In practice, the spatial random fields are often observed indirectly via some noise, blurring, treatment or covariate effects. Let the available data consist of values $y_1, \ldots, y_n$ at respective sites or sampling stations $s_1, \ldots, s_n$. Here each $s_i$ represents a small region (relative to the scale of sampling) and is often referenced by a point $(u_i,v_i)$ in $\mathbb{R}^2.$ In model based geostatistics \citep{digg:tawn:1998,digg:ribe:2007}, it is assumed that the observed data values are realizations of an explicitly specified stochastic model, such as the linear mixed model
\begin{equation}\label{eqn:geostatmodel}
  y_i = \mu + \psi(u_i,v_i) + \epsilon_i,
\end{equation}
where $\mu$ is the overall mean, $\psi$ is the underlying fractional Gaussian field \eqref{eqn:imaternspde} with the variogram function given by equation \eqref{eqn:variogPower}, and,  independent of $\psi$, random errors $\epsilon_1, \ldots, \epsilon_n$ are iid $N(0,\tau^{-1})$ residual or nugget components. The parameter $\tau^{-1}$ is popularly known as the nugget variance. 
Under the intrinsic assumption, the joint distribution of the contrasts observations $y_1,y_2,\ldots,y_n$ are then used for estimating the mean and spatial parameters, and conditional distribution of $\psi(u,v)$ given observed data values $y_1,\ldots,y_n$ is used to make predictions at an unsampled locations $(u,v) \in \mathbb{R}^2.$

For a suitable value of $m$, we next assume that the sampling stations $s_i$, $1\leq i \leq n$, can be embedded in the sublattice $\mathbb{Z}_m^2$. Furthermore, on the sublattice $\mathbb{Z}_m^2$, let the point $(u_{i,m}, v_{i,m})$ best represents the sampling station $s_i$.  We can then consider a  lattice approximation of the linear mixed model \eqref{eqn:geostatmodel} by replacing $\psi$ with $\psi^{(m)}$. This leads to an approximate model
\begin{equation}\label{eqn:latticemodel}
  y_i = \mu^{(m)} + \psi^{(m)}(u_{i,m} , v_{i,m})+ \epsilon^{(m)}_i.
\end{equation}
In the above $\mu^{(m)}$ is now the overall mean, and random errors $\epsilon_1^{(m)}, \ldots, \epsilon_n^{(m)}$ are independent of $\psi^{(m)}$ and are iid $N(0,\tau^{-1}_m)$. The nugget variance  $\tau^{-1}_m$  is analogous to $\tau^{-1}$.

One important aspect of model based geostatistics is that it explicitly describes the joint distribution of the observations, thus providing a likelihood for the parameters. It provides a complete approach to inference based on variograms which is primarily used by practitioners. For more detail on model based geostatistics we refer the readers to \citet{digg:ribe:2007} and subsequent references.

\subsection{Maximum likelihood estimation}\label{sec:MLestimation}
Generally, the Gaussian linear mixed models allow maximum likelihood methods for estimating spatial parameters of interest thus facilitating model selection via information criteria, statistical inference, and more importantly, assessment of the uncertainty of the parameter estimates. However, maximum likelihood estimation for models \eqref{eqn:geostatmodel}  and \eqref{eqn:latticemodel} presents significant challenges. In particular, exact MLE calculations for \eqref{eqn:geostatmodel} can be very challenging for any values of $\nu \ge 2$. Furthermore, $\nu \le 1$, fractional Gaussian fields are not defined pointwise but only in a distributional sense. This also presents additional complications. For MLE calculations with $\nu=1$, we refer to  McCullagh and Clifford (2006) and Dutta and Mondal (2015). Here, for ease of exposition, we restrict our discussion to $1< \nu <2$. When $1< \nu <2$, fractional Gaussian fields have stationary increments. Thus, for this range of the fractional parameter,  the marginal variances of the observations are infinite but all contrasts possess valid joint distribution. Moreover, in this case, the expected value of any contrast of the vector ${y} = (y_1,\ldots,y_n)^\top$ is zero. In the next subsection, we use these properties to advance MLE calculations.

\subsubsection{MLE for fractional Gaussian fields }\label{sec:MLiMatern}

We assume $1< \nu <2$. In the continuum model (8), the observations themselves do not possess a regular joint distribution because the marginal variances are not finite. However, in this case, all contrasts of the observations admit a non-singular multivariate normal distribution. To that end, suppose ${C}$ is an $(n-1)\times n$ matrix of orthogonal contrasts so that ${C1_n=0}$ and ${CC}^\top = {I}_{n-1},$ where $1_n$ is the $n\times 1$ vector of ones. Then the joint distribution of ${Cy}$ is multivariate normal with zero mean vector and covariance matrix 
${C}\Sigma{C}^\top + \tau^{-1}{I}_{n-1},$ where the $(i,j)$th entry of $\Sigma$ arises from the variogram \eqref{eqn:variogPower} and is given by
\[\sigma_{ij} = \dfrac{\sigma^2\pi\Gamma(\nu-\fhalf)}{16\sqrt{\alpha(\half-\alpha)}\Gamma(\nu)\Gamma(2\nu-1)\sin  (\nu\pi)}\left(\frac{(u_{i}-u_{j})^2}{4\alpha} + \frac{(v_{i}-v_{j})^2}{4(\fhalf-\alpha)}\right)^{\nu-1},\]
where $\sigma^2 > 0,$ $\tau >0,$ $1 < \nu < 2$ and $0 < \alpha < \fhalf.$ Note that, although $\Sigma$ is not non-negative definite, $C\Sigma C^\top$ is positive semi-definite. Consequently, the log-likelihood of the parameter $\theta = (\tau,\sigma^2,\nu,\alpha)$ is given by
\begin{equation}\label{eqn:loglikMatern}
2\ell(\theta) = -(n-1)\log(2\pi) - \log\det({C}\Sigma{C}^\top + \tau^{-1}{I}_{n-1}) - {y}^\top{C}^\top({C}\Sigma{C}^\top + \tau^{-1}{I}_{n-1})^{-1}{Cy}.
\end{equation}
This likelihood function is invariant to the choice of the orthogonal contrast matrix ${C}$ because different choices change the log-likelihood by an additive constant that does not depend on the parameters. ML estimates of the parameters are obtained by maximizing $\ell$ within the domain. Because the parameters $\nu$ and $\alpha$ are constrained inside intervals, the limited memory Broyden–Fletcher–Goldfarb–Shanno algorithm with box constraints (L-BFGS-B) provides a practically useful tool for ML estimation. We also obtain the numerical hessian matrix as a byproduct of the algorithm and compute the standard errors of the parameters as the square roots of the diagonals of the inverse hessian matrix.

There are some practical drawbacks of estimating the parameters using this method. First, the method requires inversion of an $(n-1)\times (n-1)$ covariance matrix, which is typically done using the dense Cholesky factorization, that requires $O(n^2)$ storage space in memory and has $O(n^3)$ computational complexity. Thus the method is only useful for moderate sample sizes. Second, the log-likelihood is not a concave function. Thus we cannot guarantee a global maximum. At the same time, maximization can run into boundary problems, meaning that maximum value is susceptible to occur at the boundary of the parameter space. 

Finally, for $\nu \ge 2$, MLE calculations get exceedingly difficult, as we need to consider a different contrast matrix $C$ that can generate all first-order increments of the observed data.


\subsubsection{MLE with lattice approximations}\label{sec:MLfracdiff}

Exact MLE calculations for the model \eqref{eqn:latticemodel} also presents challenges.  This is because unlike \eqref{eqn:variogPower}, variogram calculations \eqref{eqn:variogsddif} require expensive numerical computation. However, on any finite regular lattice, \eqref{eqn:fld} provides another alternative way to approximate the model \eqref{eqn:latticemodel}. To that end, suppose that for a specific value of $m,$ the spatial domain is embedded in a finite regular rectangular array with $r$ rows and $c$ columns (both of which depend on $m$). Then, under a restriction of $\Delta_m$ to the finite $r\times c$ array, a solution $\varphi$ to \eqref{eqn:fld} has a precision matrix $\lambda_m R^\nu$ where $\lambda_m = m^2/\sigma_m^2$ and $R$ is the $rc\times rc$ matrix representing the restriction of $\Delta_m$ to the finite $r\times c$ array. Under a column major ordering of the entries of the $r\times c$ array, \citet{dutt:mond:2015}, have shown that the $rc\times rc$ matrix $R$ admits a spectral decomposition given by
\[R  = {P}^\top \bigl(4\alpha_m {D}_{01} + 4(\fhalf-\alpha_m){D}_{10}\bigr){P},\]
with $P = P_c\otimes P_r,$ ${D}_{10} = {I}_c\otimes {D}_r$ and ${D}_{10} = {D}_c\otimes {I}_r,$ where for $l=r$ or $c,$ $P_l$ is the $l\times l$  orthogonal matrix with $(i,j)$th entry given by
\[ p_{1,j} = l^{-\shalf},\quad p_{i,j} = (2/l)^{-\shalf}\cos \bigl\{\pi(i-1)(j-\shalf)/l \big\}, \quad  i=2,\ldots,l, \quad  j=1,\ldots,l,\] 
and ${D}_l$ is the $l\times l$ diagonal matrix with $i$th diagonal entry
\[d_{i} = \sin^2\bigl\{\pi(i-1)/(2l)\bigr\},\quad  1\leq i \leq l.\]
Consequently, suppressing $m$, we revise \eqref{eqn:latticemodel} using $\varphi$ as
\[
{y} = \mu + {F}\varphi + {\varepsilon}
\]
where ${F}$ is the $n\times rc$ incidence matrix with $i$th row $f_i$ such that $f_i^\top\varphi$ gives the $\varphi-$values at $(u_{im},v_{im}),$ ${\varepsilon} = ({\varepsilon}_1^{(m)},\ldots,{\varepsilon}_n^{(m)})^\top,$ and the improper density for $\varphi$ is given by,
\begin{equation}\label{eqn:densitypsi}
 f(\varphi) \varpropto \left|\lambda_m R^\nu \right|_{+}^{\fhalf} \exp\left(-\half\lambda_m\varphi^\top{R}^\nu\varphi\right).
\end{equation}
In the above, we interpret the fractional power of $R$ via its spectral density,
\begin{equation}\label{eqn:sddifprecision}
R^\nu ={P}^\top \bigl(4\alpha_m {D}_{01} + 4(\fhalf-\alpha_m){D}_{10}\bigr)^\nu{P}.
\end{equation}

In order to estimate the parameters $\theta_m = (\tau_m,\lambda_m,\nu,\alpha_m)$, \citet{dutt:mond:2016a} takes an h-likelihood approach. Unlike the method described in Section \ref{sec:MLiMatern}, the above finite regular lattice approximations and the h-likelihood method are valid for all $\nu>0.$ The h-likelihood method goes as follows. Let ${B}$ denote the last $rc-1$ rows of the matrix ${M}$ so that ${B}\varphi$ is an $rc-1$ variate normal random vector with diagonal precision matrix ${G}$ consisting of the $rc-1$ non-zero eigen values of $\lambda_m{R}^\nu.$  Next, define the following matrices and vectors
\[{X} = 
\begin{pmatrix}  {1}_n & {F} \\ {0} & {B}\end{pmatrix},~ 
{z}=\begin{pmatrix} {y} \\ {0}\end{pmatrix},~
\bbeta = \begin{pmatrix}\mu^{(m)} \\ \varphi \end{pmatrix},~
{Q} = \begin{pmatrix} \tau_m{I}_n & {0}\\ {0} & {G}\end{pmatrix} \textrm{ and }{H=X(X^\top QX)^{\mathrm{-1}}X^\top Q}.
\]
\citet{dutt:mond:2016a} then obtain the residual likelihood (REML) function $\ell_R$ given by
\begin{equation}\label{eqn:remlhlik}
 2\ell_R(\tilde\theta) = \log\det {Q} - \log|X^\top QX|_+ - {(z-X\widehat{\bbeta})^\top Q (z-X\widehat{\bbeta})}
\end{equation}
where $\widehat{\bbeta}$ is the solution to
\begin{equation}\label{eqn:blup}
 ({X^\top QX})\bbeta = {X^\top Qz}.
\end{equation}
Traditional maximization of the log REML function uses  score equations which are obtained by equating the gradient of $\ell_R$ to zero. Thus suppose ${Q}_{1} = \partial {Q}/\partial \tau_m,$ ${Q}_{2} = \partial {Q}/\partial \lambda_m$,  ${Q}_{3} = \partial {Q}/\partial \nu,$ and ${Q}_{4} = \partial {Q}/\partial \alpha_m,$.  The score equations  that maximize the log--REML function in \eqref{eqn:remlhlik} are then given by
\[
 \half\Tr \left({Q}^{-1}{Q}_{i}\right) - \half\Tr~\bigl\{ ({X}^{\top} {QX})^{-1}{X}^{\top} {Q}_{i}{X}\bigr\} - \half({z-X}\widehat{\beta})^{\top}{Q}_{i}({z-X}\widehat{\beta}) = 0 
\]
for $i=1,\ldots,4$. Note that these score equations can also be expressed succinctly as
\begin{equation}\label{eqn:scorehlik}
\half\Tr~{(I-H)Q}^{-1}{Q}_{i} - \half({z-X}\widehat{\beta})^{\top}{Q}_{i}({z-X}\widehat{\beta}) = 0,\quad i=1,\ldots,4.
\end{equation}
Typically, Fisher's scoring method is used to solve the score equations and to obtain REML estimates. However, this also requires computation of the second derivatives  of the log REML function or the information matrix ${\mathfrak{I}}$ whose $(i,j)$th entry is equal to  
\begin{equation}\label{eqn:informationhlik}
 \mathfrak{I}(i,j) = \half\Tr\bigl\{{(I-H)Q}^{-1}{Q}_{i}{(I-H)Q}^{-1}{Q}_{j}\bigr\},
\end{equation}
which can also be used to derive standard errors of the estimates. However, computing the trace terms either in \eqref{eqn:scorehlik} or \eqref{eqn:informationhlik} are not straightforward as they require computing the diagonal entries of the hat-matrix ${H}.$ For large values of $n,$ an exact computation of these trace terms has $O(n^3)$ computation complexity and requires $O(n^2)$ memory storage space. As a practical alternative, \citet{dutt:mond:2016a} then suggests instead solving the unbiased system of equations
\begin{equation}\label{eqn:approxScoreEq}
 g_{i}(\theta_m) = \frac{1}{2K} \sum_{t=1}^{K}{u}_{t}^{\top}{Q}^{-1}{Q}_{i}({I}-{H}){u}_{t} - \half({z-X}\widehat{\beta})^{\top}{Q}_{i}({z-X}\widehat{\beta}) = 0,
\end{equation}
where ${u}_t$'s are i.i.d Rademacher random vectors with entries $\pm 1$ with probability $\fhalf$ each. Here the number of Rademacher vectors, $K,$ should be large. However, the results of \citet{dutt:mond:2016a} suggests $K=50$ retain sufficient statistical efficiency of the estimates.

\citet{dutt:mond:2016a} provide a sophisticated matrix-free trust-region algorithm for solving \eqref{eqn:approxScoreEq} that crucially depend on the matrix-free discrete cosine transformation for computing matrix-vector multiplications of the form $Pv$ and the matrix-free inverse discrete cosine transformation for computing $P^\top v$ for $v\in \mathbb{R}^{rc},$ \citep{rao:yip:1990,frig:john:2005} and a matrix-free preconditioned Lanczos algorithm for solving large system of linear equation \eqref{eqn:blup} and those involved in \eqref{eqn:approxScoreEq}.  Furthermore, this computational framework yields the standard errors of the parameter estimates as well as the best linear unbiased predictions of the random field $\varphi$ that serves as the kriged surface of the random field. Overall, in contrast to the dense-matrix computations the computational complexity of the matrix-free algorithms is essentially $O(n(\log n)^2)$ using only $O(n)$ storage in memory.


%
%

\section{Simulation studies}\label{sec:sim}

We perform two simulation studies. The goal of the first simulation study is to derive  the estimates for the fractionally differenced random field model when the data is generated from a continuum fractional Gaussian field plus a nugget effect, and to compare these estimates with the actual maximum likelihood estimates. The goal of the second simulation study is to demonstrate the scalability of statistical computation for fractional Laplacian differencing.

\subsection{An experiment with power-law variogram}\label{sec:sim1}

\begin{figure}
 \centering\includegraphics[width=0.8\textwidth]{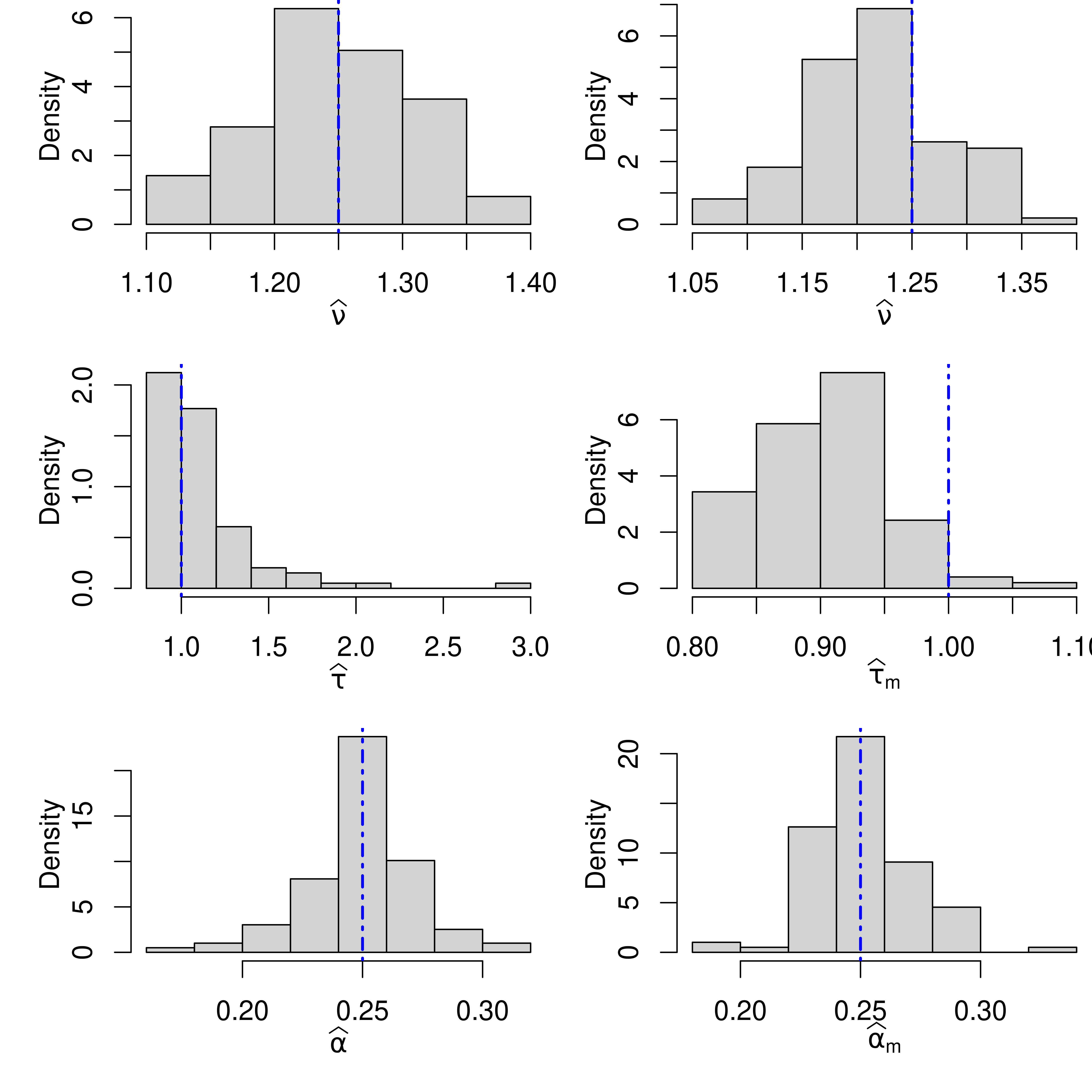}
 \caption{Histograms of the estimates of $\nu$ (top), nugget precision (middle) and anisotropy parameters (bottom) using direct ML estimation of intrinsic Mat\'ern model (left column) and h-likelihood method on the lattice model (right column).}
 \label{fig:histAccuracy}
\end{figure}

\begin{table}
 \caption{Coverage probabilities and mean widths of 95\% confidence intervals based on normal approximations.}
 \label{tab:confintAccuracy}
 \begin{center}
\begin{tabular}{|c |c c |c c |c c|} \hline
 & \multicolumn{2}{c|}{$\nu$} & \multicolumn{2}{c|}{$\log\tau~\vdots~\log\tau_m$} & \multicolumn{2}{c|}{$\alpha~\vdots~\alpha_m$} \\
 Model &  Coverage  & Width & Coverage & Width & Coverage & Width \\ \hline
Fractional Gaussian  &  100 &  1.31 & 92.9 & 0.89 & 100 & 0.19\\
fields &&&&&&\\ \hline
Fractional Laplacian  & 96 & 0.27& 60.6 & 0.24 & 94  & 0.09 \\ 
differening &&&&&&\\
\hline
\end{tabular}
 \end{center}
\end{table}

We generate data on 4000 randomly selected grid points in a 100x100 lattice embedding the unit square from an intrinsic Mat\'ern random field with $\nu = 1.25, \tau = 1,$ $\sigma^2 = 2,$ and $\alpha = 0.25.$ We compute the estimates of $\nu,\tau,\sigma^2$ and $\alpha$ using the method described in Section \ref{sec:MLiMatern}. Next, we fit the lattice model \eqref{eqn:latticemodel} the original $100\times 100$ array (so that $m=1$) and compute estimates of $\nu,$ $\tau_m,$ $\lambda_m$ and $\alpha_m.$ We repeat this process 100 times. Overall, the h-likelihood method was between 40--80 times faster than the direct ML estimation of intrinsic Mat\'ern model and in one of these simulations the direct ML method failed to converge, yielding estimates on the boundary. We discard this case from our analysis. It is expected that the analysis on the original scale with the fractionally differenced model would yield biased estimate of $\tau_m$ because it compensates or absorbs the difference between the lattice variogram and the continuum variogram as seen in Figure \ref{fig:variodiff}. Figure \ref{fig:histAccuracy} shows the histograms of these estimates from the two models along with the true values. These plots show that the lattice based fractionally differenced model provide practically useful estimates of $\nu$ and the anisotropy parameter. However, it over estimates the nugget variance (underestimates nugget precision). On the other hand, the confidence intervals and their average widths in Table \ref{tab:confintAccuracy} show that the fractionally differenced model provides shorter and more practically meaningful confidence intervals for $\nu$ and the anisotropy parameters.

\begin{figure}
 \begin{center}
  \includegraphics[width=0.9\textwidth]{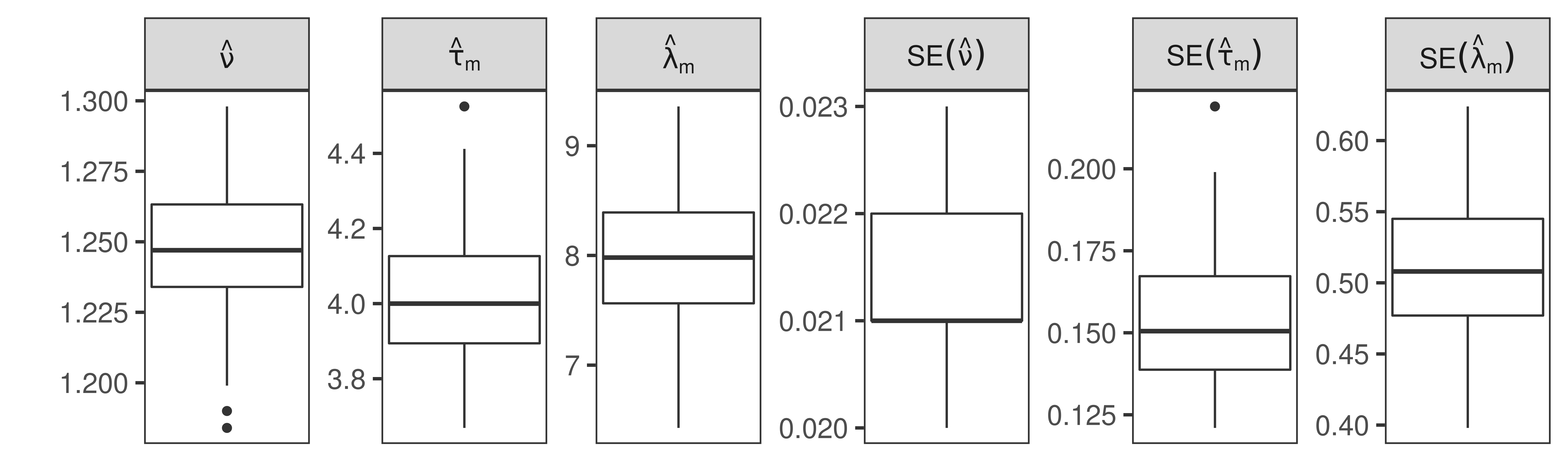}
  
  \includegraphics[width=0.9\textwidth]{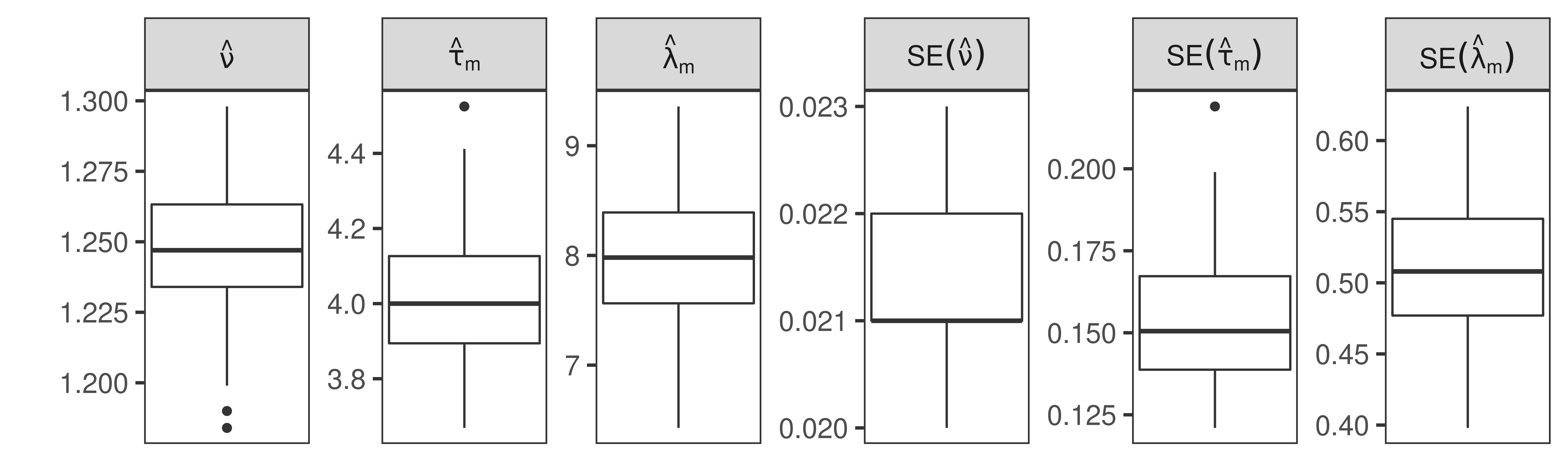}
 \end{center}
\caption{Boxplots of the parameter estimates and their standard errors. True values of the parameters are:  $\tau_m=4$ and $\lambda_m=8$ and  $\nu=1.25$ in top panel and $\nu=1.5$ in bottom panel.}
\label{fig:largesim}
\end{figure}

\subsection{Large scale computation with lattice approximations}

In this section, we demonstrate the scalability of the likelihood computations using the fractionally differenced model. To that end, we now generate data on grid points of a $256\times 256$ regular rectangular array from two fractionally differenced models. We keep $\tau_m = 4,$ $\lambda_m=8$ and fix $\alpha_m=0.25$ and use two different values $\nu=1.25$ and $\nu=1.5.$ We randomly keep 60\% of the observations resulting in a sample of size around $39321\pm125$ (mean $\pm$ sd). The data is generated from the fractionally differenced model because such the method for generating from the intrinsic Mat\'ern model runs out of the memory. Similarly, the method for fitting the intrinsic Mat\'ern model using dense-matrix computations also fail on such large datasets. In contrast, the fractionally differenced model fits without any issue on a standard personal computer. The process is repeated 100 times for each choices of $\nu$ and the resulting boxplots of the estimates and their standard errors are shown in Figure \ref{fig:largesim}. We find that the estimates are very close to the true values of the parameters and are also unbiased. Furthermore, the standard errors are also very small suggesting the estimators are statistically efficient, a fact that is also noted in \citet{dutt:mond:2016a}.

\section{Indian Ocean surface temperature from Argo floats} \label{sec:data}
\begin{figure}[htp]
 \centering\includegraphics[width=0.95\textwidth]{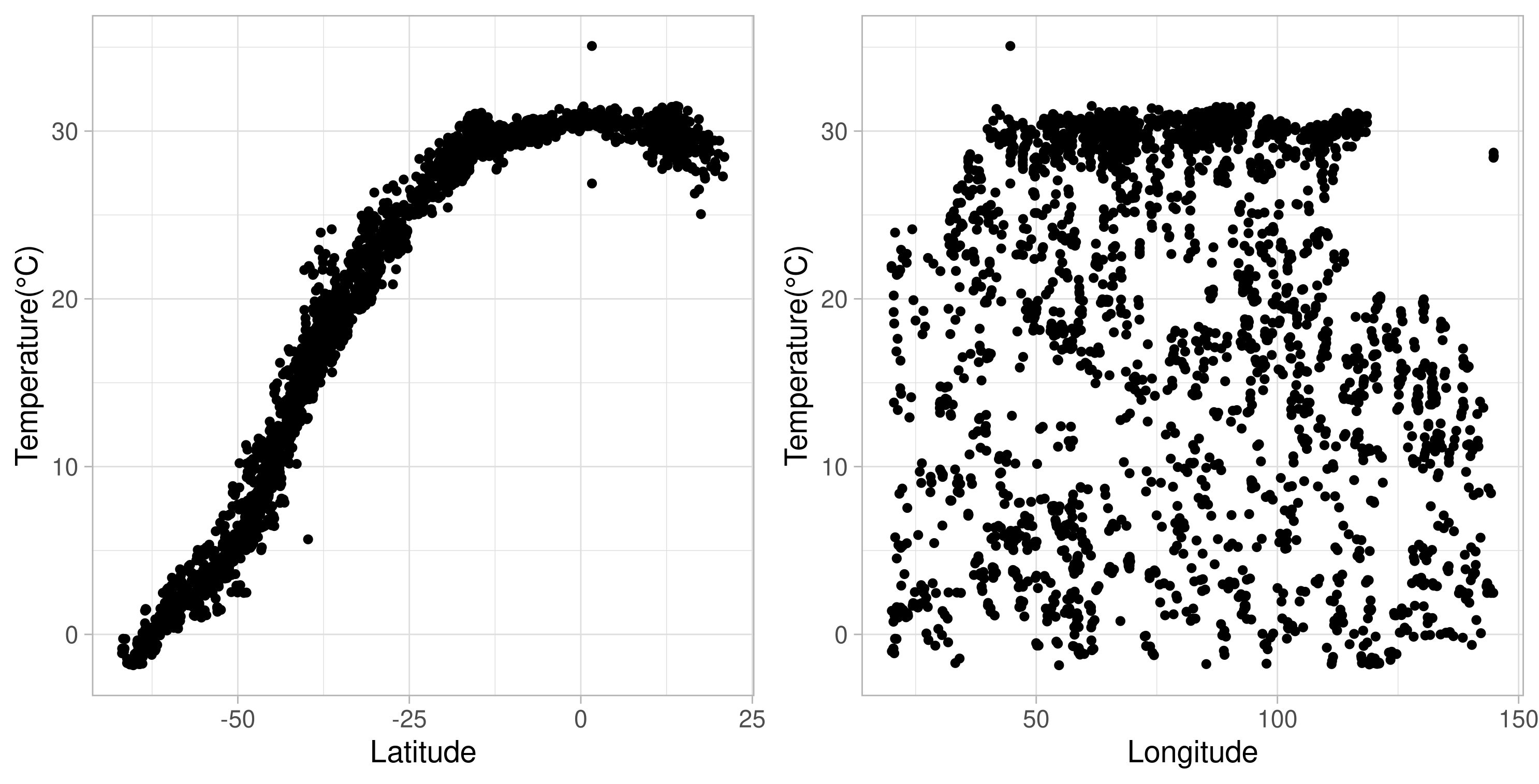}
 \centering\includegraphics[width=0.95\textwidth]{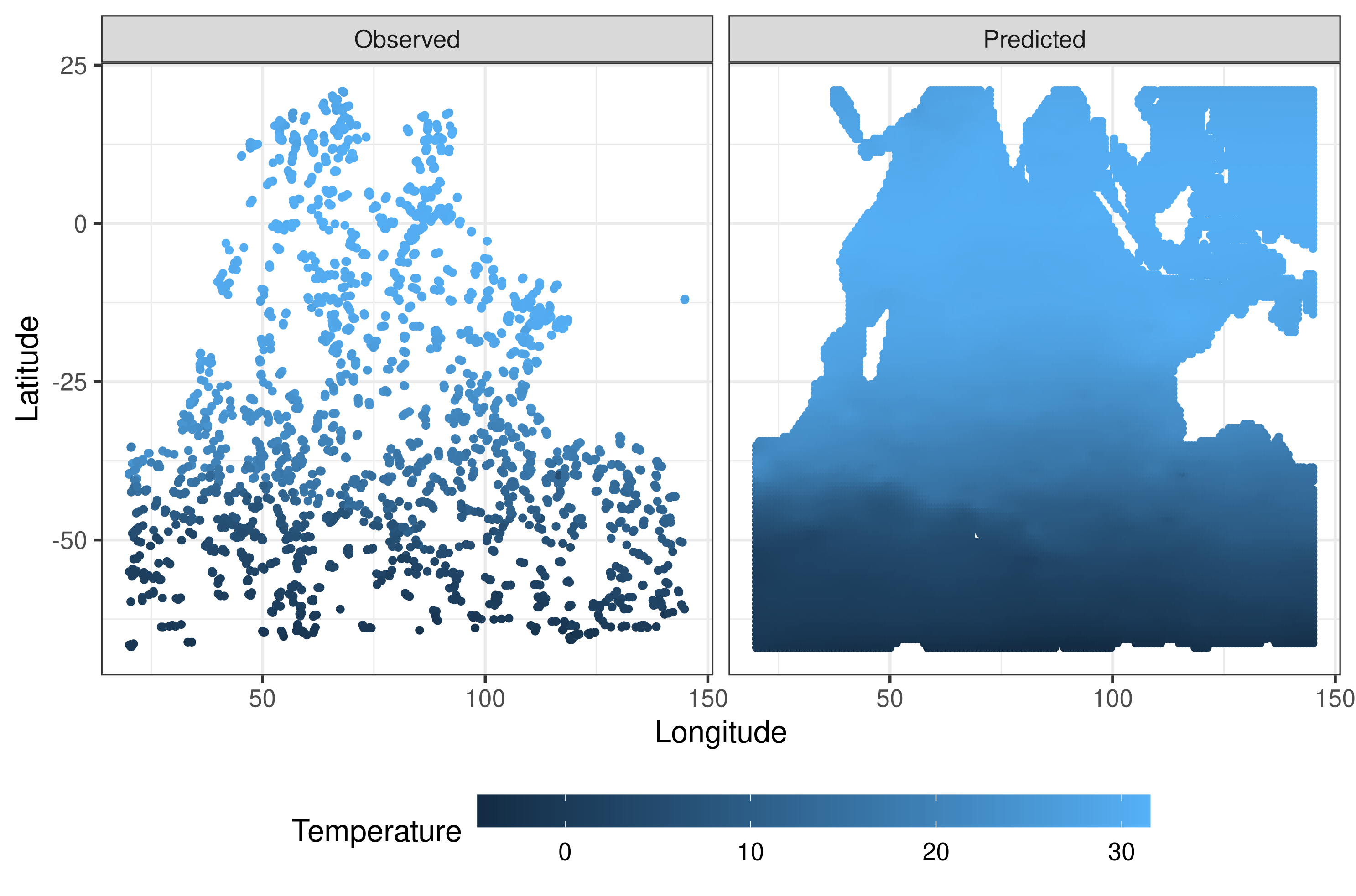}
 \caption{Sea-surface temperature (in \degc) measured by the Argo floats in the Indian Ocean during April, 2020. Top: Observed temperature against the geographic coordinates. Bottom left: Image of observed temperature values and Right: Krigged sea-surface temperature.}
 \label{fig:argodata}
\end{figure}

The Argo Program is part of the Global Ocean Observing System from an international collaboration among more than 30 countries from all continents that provides useful data on important ocean variables. Conceived in the early 2000s, the Argo fleet now consists of more than 4000 drifting battery-powered machines called Argo \emph{floats} that are deployed worldwide. These floats weigh around 20-30kg each, and typically probe the drifts at a depth where they are stabilized by their buoyant (around 1km). Every 10 days or so, these floats change their buoyant and dive to a depth of 2km and then rise to the water-surface measuring conductivity and temperature profiles as well as pressure, over about 6 hours. From the surface, they transmit their location as well as the collected data to satellites and dives back to their drifting depth. In this section, we analyze the monthly data on sea-surface temperature in the Indian Ocean obtained from April 1, 2020 till April 30, 2020.  These data were collected and made freely available by the International Argo Program and the national programs that contribute to it  \citep{argo2020}.   After removing the erroneous measurements as described on the Argo website, we obtain 2525 observations of sea-surface temperature (in \degc) and plot them in the bottom left panel of Figure \ref{fig:argodata}. Note that the Argo floats are quite scattered over the Indian Ocean and the temperature are clearly spatially auto-correlated. Furthermore, from the top panel of Figure \ref{fig:argodata} we can see that the temperature variation seems to be more along the latitude compared to the longitude, as one would naturally expect. In fact, this suggests that an (intrinsically) stationary model may not be accurate. To account for this trend along the latitude, we fit a quadratic mean model
\begin{equation}\label{eqn:quadratictrend}
\mu(l) = a_0 + a_1 l + a_2 l^2    
\end{equation}
where $l$ denotes the latitude using ordinary least squares. Next, we obtain the residuals from this quadratic mean model and use them as response for the following spatial analysis.

\begin{figure}[htp]
 \centering\includegraphics[width=0.6\textwidth]{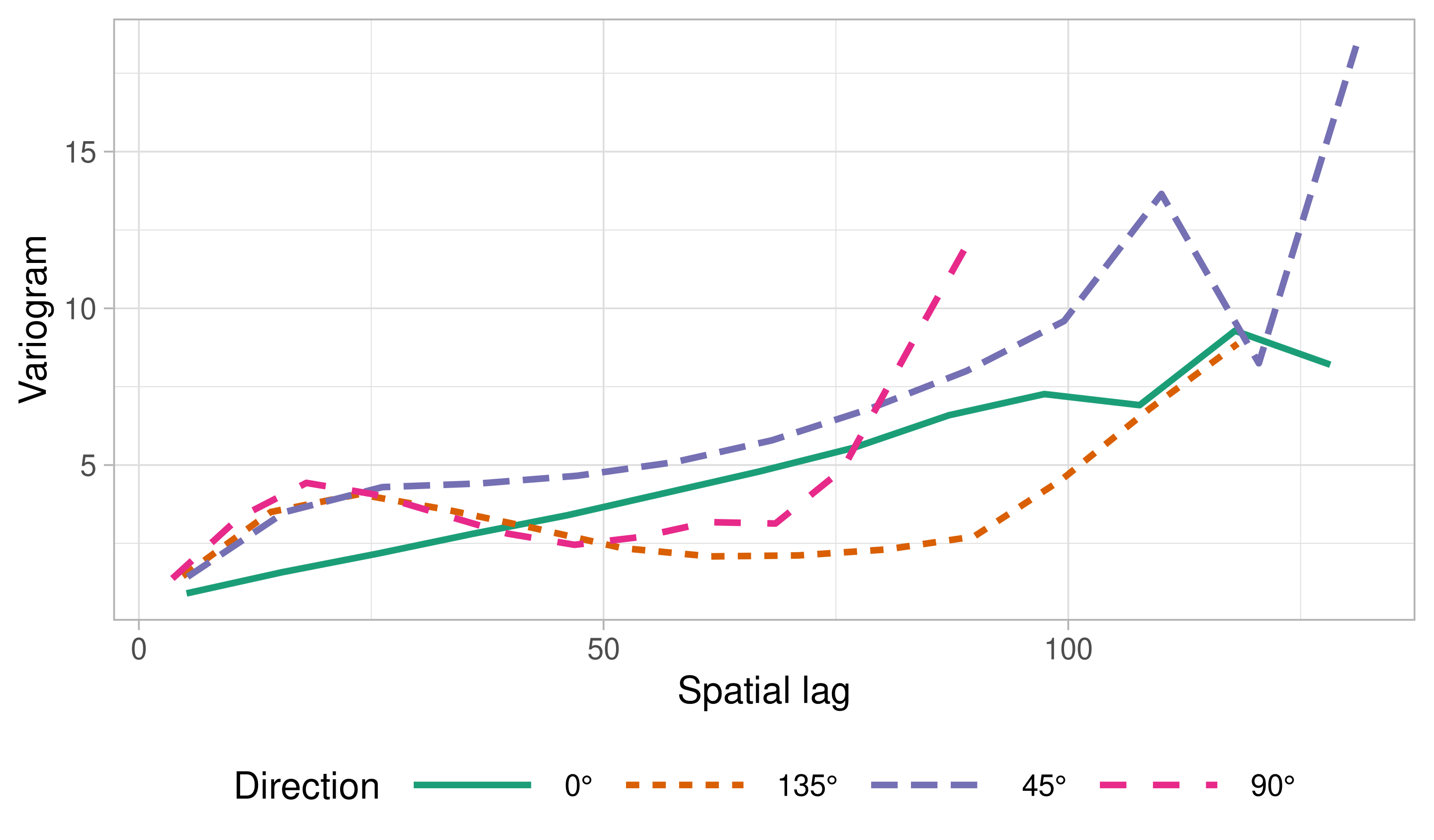}
 \caption{Directional variograms of the residual temperature values.}
 \label{fig:argo-variogram}
\end{figure}

First we run some exploratory analyses. We compute and plot the empirical variogram using the R package \textsf{geoR} along the four directions and plot it using the R package \textsf{ggplot2}.
\begin{verbatim}
library(geoR)
library(dplyr)
library(ggplot2)
library(RColorBrewer)

temp = read.table("april-data.txt")
names(temp) = c("Latitude","Longitude","Temperature","Resid.quad")
temp.geo = as.geodata(temp,data.col = 4) 
# Using the residuals from quadratic model as response
vg4 = variog4(temp.geo)
vg.gg = data.frame(h = vg4$`0`$u , v = vg4$`0`$v, Direction='0°') %>% 
  rbind(., data.frame(h = vg4$`45`$u , v = vg4$`45`$v, Direction='45°')) %>%
  rbind(., data.frame(h = vg4$`90`$u , v = vg4$`90`$v, Direction='90°')) %>%
  rbind(., data.frame(h = vg4$`135`$u , v = vg4$`135`$v, Direction='135°'))

ggplot(vg.gg,aes(x=h,y=v,group=Direction,color=Direction,lty=Direction)) + 
  geom_line(size=1.5) + xlab("Spatial lag") + ylab("Variogram") + 
  theme_light(base_size = 16) + scale_color_brewer(palette="Dark2") +
  theme(legend.position = "bottom",legend.key.width = unit(2.5,"cm"))
\end{verbatim}
The dataset is also available by an email request to the corresponding author. The directional variograms are shown in Figure \ref{fig:argo-variogram}. We see that the variogram increases more along the 90\deg and the 45\deg directions supporting that there is more spatial variability across the latitude. Furthermore, the variograms along these directions do not seem to reach a sill, suggesting that an intrinsic model could be more appropriate for the data.

\begin{table}[t]
\caption{Estimates of the spatial parameters from the spatial linear mixed model based on fractional Gaussian field (FGF) and its lattice approximation (FLD). Standard errors are shown in parentheses.}
\label{tab:argoestimates}
 \begin{center}
  \begin{tabular}{cccc} \hline 
   Model & $\widehat{\nu}$ & $\widehat\alpha~\vdots~\widehat{\alpha}_m$ & $\widehat{\tau}~\vdots~\widehat{\tau}_m (\degc^{-2})$  \\ \hline 
   FGF & 1.400 ( 0.114 ) &  0.058 ( 0.0314 ) &  3.59 ( 0.285 ) \\ 
   FLD & 1.426 ( 0.051 ) &  0.074 ( 0.012 ) &  3.11 ( 0.267 ) \\
   \hline
  \end{tabular}

 \end{center}

\end{table}

We first fit the intrinsic Mat\'ern model to the data using the method described in Section \ref{sec:MLiMatern}. The estimates of $\nu,\alpha$ and $\tau$ are shown in the first row of Table \ref{tab:argoestimates} and the estimate of $\sigma^{-2}$ is $\widehat{\sigma}^{-2} = 8.987\degc^{-2}$ (s.e. $1.19\degc^{-2}$). The estimate of $\alpha$ corroborates the observation that the temperature varies more across the latitude than the longitude.

Next, we fit the fractionally differenced process to the data using the method described in Section \ref{sec:MLfracdiff}. To that end, we embed the region bounded between by 21\deg to the North, 67\deg to the South, 20\deg to the West and 145\deg to the East in a  $128\times180$ regular rectangular array so that each pixel is approximately $0.6875\deg$ latitude by $0.694\deg$ longitude. Next we average the residuals from the quadratic model falling inside the same lattice pixel, resulting in around 7.43\% observed pixels. We use 50 Rademacher variables for stochastically approximate the score equations. Abusing the notation, we drop the subscript $m$ from $\tau_m,$ $\alpha_m$ and $\lambda_m,$ as $m$ is implicitly chosen via the array dimensions. The estimates of $\nu,\alpha$ and $\tau$ are shown in the second row of Table \ref{tab:argoestimates} and the estimate of $\lambda$ is $\widehat{\lambda} = 15.381\degc^{-2}$ (s.e. $3.35\degc^{-2}$). The results largely agree with the findings in Section \ref{sec:sim1}. In particular,  although the estimates of $\nu$ and the anisotropy parameters are very close from the two methods, the fractionally differenced model yields smaller standard errors of these estimates. The slight discrepancy in the estimates of $\alpha$ occurs because the pixels are not exact squares. On the other hand, the estimate of the nugget precision is lower from the fractionally differenced model than the intrinsic Mat\'ern model.

Note that as a byproduct of fitting the fractionally differenced model, we also obtain the best linear prediction $\widehat{\varphi}$ of the underlying spatial random field $\varphi$. The image $\widehat{\varphi}$ plus the quadratic mean from \eqref{eqn:quadratictrend} is shown on the right panel of Figure \ref{fig:argodata}. Note that the fine scale features of the temperature gradient are more prominent in the krigged map. Such interpolated maps are often useful in studying other oceanic and atmospheric activities.

\section{Concluding remarks}\label{sec:discussion}

This book chapter  presents a brief review on fractional Gaussian fields, their lattice based approximations, and connections to the intrinsic and stationary M\'atern, power-law and other generalized random fields. Likelihood based inference methods have been developed for spatial linear mixed models based on the fractional Gaussian random fields and fractional Laplacian differencing on regular lattice. Computational methods for maximum likelihood estimation of parameters have been described and compared. Using both simulation and data examples, it is demonstrated that the lattice based model facilitate faster and more stable statistical computations of the maximum likelihood estimators than the model based on fractional Gaussian fields, while providing practically close estimates of dependence and anisotropy parameters. Moreover, the h-likelihood method for the model on regular lattice provide more useful estimates of uncertainty and confidence intervals for the aforementioned parameters than the maximum likelihood method for the geostatistical model based on the fractional Gaussian fields.

It must be stressed that there are definite advantages in discretizing the space using a regular lattice instead of other well known ideas such as triangulation \citep{lind:rue:2011} and neighborhood selection \citep{datt:bane:2016} of irregularly distributed sampling stations. One advantage is the explicit spectral decomposition that allows for the use of fractional values $\nu$ and provides fast matrix-free computation in terms of discrete cosine transformation. Another advantage is accommodation of geometric anisotropies. It must be noted that irregular discretizations do not permit us  to accommodate geometric anisotropies in any obvious way.

Our presentation has focused on fractional models with long range dependence. To study short range dependence, we can consider stationary Mat\'ern covariance models \citep{hask:cull:2007,stei:2012,guin:mont:2017}. As fractional Gaussian fields are limiting cases of Mat\'ern models,  we can also obtain lattice approximations of the latter. Under the same setup as Section \ref{sec:MLfracdiff}, the inverse variance covariance matrix of this approximate Mat\'ern model takes the form
\begin{equation}\label{eqn:stationarycase}
\lambda_mR^\nu = \lambda_m P^\top (\kappa_{m} + 4\alpha_m D_{01} + 4\alpha'_{m} D_{10})^\nu P
\end{equation}
where $\kappa_m,\alpha_m,\alpha'_m$ are non-negative and $\kappa_m + 4\alpha_m + 4\alpha'_m = 2.$
Thus, both Mat\'ern models and their lattice approximations contain an additional range parameter, and at first it may appear that Mat\'ern models and their lattice approximations have added flexibility due to the extra range parameter. 
However, inclusion of an unknown finite range parameter often leads to long flat ridges in the likelihood function, which in turn incur substantial numerical instability in the MLE computations. This, for example, has been observed in \citep{lim:chen:2017} and also in our own experiments with the lattice approximation \eqref{eqn:stationarycase}. Interestingly, the work of \citet{zhan:2004} suggest that the scale and the range cannot both be estimated consistently. In fractional fields, we set the range parameter at infinity. In the short range dependence case, we can also fix the range parameter to a finite number to lessen numerical instabilities and to enhance interpretability. We can then proceed computation as presented in Section~3 of this chapter.

\section*{Acknowledgement} The authors thank an anonymous referee for helpful comments. Dutta's research was supported in part by the United States Department of Agriculture (USDA) National Institute of Food and Agriculture  (NIFA)  Hatch  project  IOW03617. Mondal's research was supported by the National Science Foundation (NSF) award DMS-1916448. The content presented in this chapter are those of the authors and do not necessarily reflect the views of NIFA, USDA and NSF. 

\bibliographystyle{bathx}
\bibliography{refs.bib}

\section{Appendix}

\subsection{Matlab codes for Section \ref{sec:MLiMatern}}

\begin{verbatim}
function [x,se] = mlFGF(row,col,y,initial)
% INPUT:
% row: n x 1 vector of x-coordinates
% col: n x 1 vector of y-coordinates
% y: n x 1 vector of observations
% initial: Starting values [tau, lambda, nu, alpha] where lambda is 1/sigma^2
%
% OUTPUT:
% x: estimate of the parameters in otder [tau,lambda,nu,alpha]
%    where lambda = 1/sigma^2
% se: standard error of the parameters
n = length(row);
if n ~= length(y) || n ~= length(col)
    error('lengths of the three vectors must be equal');
end
if (initial(4) > 0.5) || (initial(4) < 0)
    error('Initial for anisotropy parameter must be between 0 and 0.5');
end
diffrow2 = (row - row').^2;
diffcol2 = (col - col').^2;
% orthogonal contrast matrix
[Cmat,~] =  qr(eye(n) - ones(n)/n);
Cmatt = Cmat(:,1:n-1);
Cmat = Cmatt';
Cy = Cmat * (y-mean(y));
f = @(logx) -loglikFGF(logx,diffrow2,diffcol2,Cmat,Cmatt,Cy,n);
logx0 = log(initial);
logx0(3) = log(initial(3)-1);
logx0(4) = log(2*initial(4)/(1-2*initial(4)));
[x, ~, ~, ~, ~, hess] = fminunc(f,logx0);
x(1:2) = exp(x(1:2));
x(3) = exp(x(3)) + 1;
se = sqrt(diag(inv(hess)));
se(1:3) = x(1:3).*se(1:3);
beta = exp(x(4));
x(4) = 0.5*beta/(1+beta);
se(4) = sqrt(beta)/(1+beta) * se(4)/2;
end

% function for computing the fractional Gaussian field log-likelihood
function v = loglikFGF(logx,diffrow2,diffcol2,Cmat,Cmatt,Cy,n)
ly = exp(logx(1));
lp = exp(logx(2));
nu = exp(logx(3)) + 1;
beta = exp(logx(4));
beta = 0.5*beta/(1+beta);
const = pi^1.5*gamma(nu-0.5)/(16*sqrt(pi) * gamma(nu)* gamma(2*nu-1))/...
        (sin(nu*pi) * sqrt(beta*(0.5-beta)));
h = diffrow2/(4*beta) + diffcol2/(4*(0.5-beta));
ucont = const * (h.^(nu-1));
Sigma = (Cmat*ucont*Cmatt)/lp + eye(n-1)/ly;
R = chol(Sigma);
z = R'\Cy;
v = -sum(log(diag(R))) - 0.5*sum(z.^2);
end
\end{verbatim}

\subsection{Matlab codes for Section \ref{sec:MLfracdiff}}

This code requires two functions \texttt{dct2mod} and \texttt{idct2mod} which takes input a matrix or vector with $mn$ entries and $m$ and $n$ and computes the discrete cosine transformation  and the inverse discrete cosine transformation of the $m\times n$ matrix using the column-major format. The returned value must be a $mn\times 1$ vector.
\begin{verbatim}
function [x, se, psi] = fracdiffML(y,w,initial,nseed)
% [x, se, psi] = iMaternRemlisotropic(y,w,initial,nseed)
% y = r x c matrix of observations
% w = r x c incidence matrix (0 = missing pixel, 1 = observed pixel)
% initial guess for [lambda_y;lambda_psi;nu; alpha]
% nseed : number of Rademacher variables (optional).
% OUTPUT:
% x: REML estimate of precision parameters [lambda_y;lambda_psi;nu]
% se: standard errors of x
% psi: BLUP of the random effects (r x c matrix)
if nargin == 3
 nseed = 50;
end
[r, c] = size(y);
q = r*c;
yield = y(w>0);
n = length(yield);
idx = find(w>0);
F = sparse(1:n,idx,1,n,q);
xr = sin(0.5*pi*(0:(r-1))'/r).^2;
xc = sin(0.5*pi*(0:(c-1))'/c).^2;
xxr = kron(ones(c,1),xr);
xxc = kron(xc,ones(r,1));
FF = full(diag(F'*F));
Z = F'*yield;
eve = rng; % backup the random number generator
rng(2441139);
RadVar = 2*(rand(n+q-1,nseed) < 0.5) - 1;
rng(eve);
options = optimset('Display','iter','TolFun',0.01,...
          'TolX',0.001,'MaxFunEvals',500,'MaxIter',40);
x0 = [log(initial(1)); log(initial(2)); log(initial(3));...
       log( 2*initial(4)/(1-2*initial(4)))];
gr = @(pars) gradfunAniso(pars,F,FF,Z,yield,xxr,xxc,r,c,n,q,RadVar,nseed);
[x, fval, exitflag, output, hess] = fsolve(gr,x0,options);
se = sqrt(diag(inv(-hess)));
ly = exp(x(1));
lp = exp(x(2));
nu = exp(x(3));
beta = 0.5/(1 + exp(-x(4)));
se(1) = se(1)*ly;
se(2) = se(2)*lp;
se(3) = se(3)*nu;
se(4) = 0.5*se(4) * exp(x(4))/(1 + exp(x(4)))^2;
x = [ly;lp;nu;beta];
L = lp*(4*beta*xxr + 4*(0.5-beta)*xxc).^nu;
PtZ = dct2mod(Z,r,c);
precon_mat = 1./(ly + L);
mxf = @(vv) precon_mat.*(ly*dct2mod(FF.*idct2mod(precon_mat.*vv,r,c),r,c) + ...
        L.*precon_mat.*vv);
[psi0 ,flag ,rel, iter] = symmlq(mxf,ly*precon_mat.*PtZ,1e-12,q);
psi1 = idct2mod(precon_mat.*psi0,r,c);
psi = reshape(psi1,r,c);
end

% Subfunction for computing the score equations.
function grad = gradfunAniso(pars,F,FF,Z,yield,xxr,xxc,r,c,n,q,RadVar,nseed)
ly = exp(pars(1));
lp = exp(pars(2));
nu = exp(pars(3));
beta = 0.5/(1 + exp(-pars(4)));
lastelt = @(v) v(2:q);
Xt = @(vv) F'*vv(1:n) + idct2mod([0;vv(n+1:n+q-1)],r,c);
X = @(vv) [F*vv; lastelt(dct2mod(vv,r,c))];
Q = [ly*ones(n,1); lp*(4*beta*xxr(2:q) + 4*(0.5-beta)*xxc(2:q)).^nu ];
dQ1 = [ones(n,1);zeros(q-1,1)];
dQ2 = [zeros(n,1);(4*beta*xxr(2:q) + 4*(0.5-beta)*xxc(2:q)).^nu];
dQ3 = [zeros(n,1);Q(n+1:n+q-1).*log(4*beta*xxr(2:q) + 4*(0.5-beta)*xxc(2:q))];
dQ4 = [zeros(n,1); 4*lp*nu*(xxr(2:q) - xxc(2:q)).*(4*beta*xxr(2:q) + ...
       4*(0.5-beta)*xxc(2:q)).^(nu-1)];
L = lp*(4*beta*xxr + 4*(0.5-beta)*xxc).^nu;
PtZ = dct2mod(Z,r,c);
precon_mat = 1./(ly + L);
mxf = @(vv) precon_mat.*(ly*dct2mod(FF.*idct2mod(precon_mat.*vv,r,c),r,c) + ... 
      L.*precon_mat.*vv);
% Compute the BLUP
[psi0 ,flag ,rel, iter] = symmlq(mxf,ly*precon_mat.*PtZ,1e-12,q);
psi = idct2mod(precon_mat.*psi0,r,c);
res2 = ( [yield; zeros(q-1,1)] - X(psi) ).^2;
g1 = 0; g2 = 0; g3 = 0; g4 = 0;
% Computing the score function
parfor t=1:nseed
    v0 = Xt(Q.*RadVar(:,t));
    v0 = dct2mod(v0,r,c);
    [v1, flag] = symmlq(mxf,precon_mat.*v0,1e-12,q);
    v = idct2mod(precon_mat.*v1,r,c);
    v = RadVar(:,t) - X(v);
    g1 = g1 + sum(RadVar(:,t).*dQ1.*v./Q)/nseed;
    g2 = g2 + sum(RadVar(:,t).*dQ2.*v./Q)/nseed;
    g3 = g3 + sum(RadVar(:,t).*dQ3.*v./Q)/nseed;
    g4 = g4 + sum(RadVar(:,t).*dQ4.*v./Q)/nseed;
end
g1 = g1 - sum(res2.*dQ1);
g2 = g2 - sum(res2.*dQ2);
g3 = g3 - sum(res2.*dQ3);
g4 = g4 - sum(res2.*dQ4);
grad = [g1;g2;g3;g4];
grad = 0.5*grad.*[ly;lp;nu;0.5*exp(pars(4))/(1 + exp(pars(4)))^2];
end
\end{verbatim}

\end{document}